\documentstyle[12pt]{article}
\begin{document}
\vspace{15mm}
\begin{center}
{\large \bf Perturbative QCD predictions for the
small $x$ behaviour}\\
{\large \bf   of unpolarized and polarized
deep inelastic scattering structure functions} \footnote{
to appear in special issue of Acta Physica Polonica\\
in honour of Professor Wojciech Kr\'olikowski's 70th birthday
}\\
\bigskip
J. Kwieci\'nski \\
Department of Physics\\
University of Durham\\
Durham, UK\\
and\\
Department of Theoretical Physics\\
H. Niewodnicza\'nski Institute of Nuclear Physics\\
Krak\'ow, Poland. \footnote{Permanent address}\\
\end{center}
\vspace{40mm}
\begin{abstract}
The perturbative QCD predictions for the small $x$ behaviour of the
nucleon
structure functions $F_{2,L}(x,Q^2)$ and $g_1(x,Q^2)$ are summarized.
The importance of the double logarithmic terms
for the small $x$ behaviour of the spin
structure function $g_1(x,Q^2)$  is emphasised. These terms correspond to the
contributions
containing  the leading powers of $\alpha_s ln(1/x)^2$ at each order
of the perturbative
expansion.
In the non-singlet case they  can be approximately accounted for by the
ladder diagrams with quark (antiquark) exchange.
We solve the corresponding integral
equation with the running coupling effects taken into account and
present  estimate of the effective slope controlling the small $x$
behaviour of the non-singlet spin structure function $g_1(x,Q^2)$ of
a nucleon.
\end{abstract}
\newpage
\section*{1. Introduction}
Analysis of the deep inelastic scattering structure functions in the limit
of small values of the Bjorken parameter $x$ allows to test
perturbative QCD in a
new and hitherto unexplored regime \cite{BCKK,ADM1,JK1}.
The advent of the HERA $ep$
collider has opened up an experimental possibility to confront the theoretical
QCD expectations  with the experimental data on deep inelastic scattering in
the small $x$ region.  The relevant theoretical framework is provided in this
case  by
the Balitzkij, Fadin, Kuraev, Lipatov (BFKL) equation for the
unintegrated gluon distribution $f(x,Q_t^2)$, where $Q_t$ denotes the
transverse momentum of the gluon and $x$ its longitudinal momentum fraction
\cite{BFKL}.
This equation sums the leading powers of $ln(1/x)$ at each order of the
perturbative expansion i.e. it corresponds to the leading
$ln(1/x)$ approximation.  The small $x$ behaviour of the structure functions
is driven by the gluon through the $g \rightarrow q \bar q$ transitions \cite
{AKMS}.\\

Perturbative QCD predicts a  strong
increase  of structure functions with the
decreasing parameter $x$ and the experimental data from HERA
are consistent with this prediction \cite{H1,ZEUS}.
This increase is much stronger than that
which would follow from the expectations based on the "soft" pomeron
exchange mechanism with the soft pomeron intercept
$\alpha_{soft} \approx 1.08$
as determined from the phenomenological analysis of total hadronic
and real photoproduction cross-sections \cite{DOLA}.\\

The small $x$ behaviour of structure functions
for fixed $Q^2$ reflects the high energy behaviour of
the
virtual Compton scattering total cross-section with increasing total CM energy
squared $W^2$ since $W^2=Q^2(1/x-1)$. The Regge pole exchange picture
 \cite{PC} is
therefore quite appropriate for the theoretical description
of this behaviour.  The high energy behaviour which follows
from perturbative QCD is often refered to as being related to the "hard"
pomeron in contrast to the soft pomeron describing the high energy
behaviour of hadronic and photoproduction cross-sections.\\

The pomeron does not contribute  to the spin dependent structure
function $g_1(x,Q^2)$ which is  controlled by the exchange
of
reggeons corresponding to axial vector mesons \cite{IOFFE}.
The pomeron also decouples, of course, from the "non-singlet"
part of the unpolarized structure functions i.e.from  the combination
$F_2^p-F_2^n$ which at small $x$ is  controlled by
the $A_2$ reggeon exchange.  The  small $x$ behaviour of
the structure function $g_1(x,Q^2)$ in perturbative QCD
is  sensitive to a new class of
double logarithmic $ln(1/x)$ contributions i.e. to the terms which
contain  powers of $\alpha_s ln^2(1/x)$ in the perturbative expansion
\cite{GORSHKOV,JK2,KL,EMR,BER}.
These terms invalidate naive Regge pole model expectations and  generate the
singular small
$x$  behaviour of the structure function $g_1(x,Q^2)$.\\

The purpose of this paper is to briefly summarize some of the
QCD expectations concerning  the small $x$ behaviour of the unpolarized and
polarized structure functions of the nucleon. In the next section we recall
the Regge pole model expectations
for the small $x$ behaviour of both the  polarized and unpolarized structure
functions while   section 3  is  devoted to the discussion
of the BFKL equation \cite{BFKL}  and its generalization based on angular
ordering, i.e. the Catani, Ciafaloni, Fiorani and Marchesini (CCFM)
equation \cite{CIAF,CCFM,KMS1}.  In Sec. 4 we discuss the small $x$ behaviour
of the polarized structure function $g_1(x,Q^2)$ concentrating
for simplicity on its non-singlet part.  We  recall the
(approximate) integral equation which resums the double logarithmic
terms and  discuss its solution after taking into account
the running coupling effects. We  present an estimate of the
effective
slope $\lambda$ controlling the small $x$ behaviour of the spin-dependent
structure function $g_1^{NS}(x,Q^2)=g_1^p(x,Q^2)-g_1^n(x,Q^2) \sim x^{-\lambda}
$.   Sec. 5 will contain our main conclusions.\\

\medskip\medskip\medskip
\section*{2. Regge pole model expectations for the small $x$ behaviour
of structure functions}
The Regge pole model describes the high energy behaviour of
scattering amplitudes in terms of the exchange of reggeons
which correspond to poles in the complex angular momentum plane in the
crossed channel \cite{PC}.
The total cross-sections are related through the optical theorem to the
imaginary
part of the scattering amplitudes in forward direction (i.e.
for $t=0$ where $t$ denotes the momentum transfer squared).  The Regge
pole model description
of the scattering amplitudes when combined with the optical
theorem gives therefore
automatically predictions for the
high energy behaviour of the total cross-sections:
\begin{equation}
\sigma^{tot}_{ab}(s) \sim {1\over s} Im A_{ab;ab}(s,t=0)= \sum_i \beta_i
\left({s\over s_0}\right)^{
\alpha_i(0)-1}.
\label{regge}
\end{equation}
The quantities $\alpha_i(0)$ are the intercepts of the Regge trajectories
and $\beta_i$  denote the couplings. \\

The Regge pole exchange is to a large extent a generalization of the single
particle exchange and so the reggeons carry the same
quantum numbers as hadrons  which particular  Regge pole trajectories
correspond to.  There is an obvious hierarchy of Regge  pole contributions
depending upon the magnitude of their intercept.  The contribution
with highest intercept corresponds to pomeron which carries
vacuum quantum numbers.\\

The high energy behaviour of the total
hadronic and (real) photoproduction cross-sections can be economically
described by two contributions: an (effective) pomeron with its
intercept slightly above unity ($\sim 1.08$) and the leading meson
Regge trajectories
with  intercept $\alpha_R(0) \approx 0.5$ \cite{DOLA}.
The reggeons can be
identified as corresponding to $\rho, \omega$, $f$ or $A_2$
exchange(s) depending
upon the quantum numbers involved. All these
reggeons have approximately the same intercept.
One refers to the pomeron obtained
from the phenomenological analysis of  hadronic  total cross
sections as the "soft" pomeron since the bulk of the processes building-up
the cross sections are  low $p_t$ (soft) processes.\\

The Regge pole model gives the following parametrization of the deep
inelastic scattering structure function $F_2(x,Q^2)$ at
small $x$
\begin{equation}
F_2(x,Q^2)=\sum_i \tilde \beta_i(Q^2) x^{1-\alpha_i(0)}.
\label{reggef}
\end{equation}
The relevant reggeons are
those which can couple to two (virtual) photons.  The (singlet) part
of the structure function $F_2$ is controlled at small $x$ by
pomeron exchange, while the non-singlet part $F_2^{NS}=
F_2^p-F_2^n$ by  $A_2$ reggeon.
Neither pomeron nor $A_2$ reggeons  couple to the
spin structure function $g_1(x,Q^2)$ which is described at small
$x$ by the  exchange of reggeons corresponding to
axial vector mesons \cite{IOFFE} i.e.to  $A_1$ exchange for the non-singlet
part $g_1^{NS} = g_1^{p}-g_1^{n}$ etc.
\begin{equation}
g_1^{NS}(x,Q^2)=\gamma(Q^2) x^{-\alpha_{A_1}(0)}
\label{gnsa1}
\end{equation}
  The  reggeons which correspond to axial vector mesons
are expected to
have very low intercept (i.e. $\alpha_{A_1} \le 0$ etc.)\\

Several  of the Regge pole model expectations
for structure functions
are modified by  perturbative QCD effects as will be briefly described
in the forthcoming Sections.
\medskip\medskip\medskip
\section*{3. BFKL pomeron and QCD predictions for the small $x$ behaviour of
the
unpolarized structure
function}
At small values of the parameter $x$ the dominant role is played by the
gluons.  The basic quantity is an unintegrated gluon distribution
$f(x,Q_t^2)$ where $x$ denotes the momentum fraction
of a parent hadron carried by a gluon and $Q_t$  its transverse
momentum.  The unintegrated distribution $f(x,Q_t^2)$
is related in the following way to the more familiar scale dependent
gluon distribution $g(x,Q^2)$:
\begin{equation}
xg(x,Q^2)=\int^{Q^2} {dQ_t^2\over Q_t^2} f(x,Q_t^2).
\label{intg}
\end{equation}
In the leading logarithmic approximation the unintegrated
distribution $f(x,Q_t^2)$ satisfies
the BFKL equation \cite{BFKL} which has the following form:
$$
f(x,Q_t^2)=f^0(x,Q_t^2)+
$$
\begin{equation}
\bar \alpha_s \int_x^1{dx^{\prime}\over
x^{\prime}} \int {d^2 q\over \pi q^2}
\left[{Q_t^2 \over (\mbox{\boldmath $q$}+
\mbox{\boldmath $Q_t$})^2}
f(x^{\prime},(\mbox{\boldmath $q$}+
\mbox{\boldmath $Q_t$})^2)-f(x^{\prime},Q_t^2)\Theta(Q_t^2-q^2)\right]
\label{bfkl}
\end{equation}
where
\begin{equation}
\bar \alpha_s={3\alpha_s\over \pi}.
\label{alphab}
\end{equation}
The first and the second
terms  in the right hand side of  eq. (\ref{bfkl}) correspond
to the real gluon emission with $q$ being the transverse
momentum of the emitted gluon, and to the virtual corrections respectively.
$f^0(x,Q_t^2)$ is a suitably defined inhomogeneous term.\\

After resumming the virtual corrections and "unresolvable"  gluon
emissions ($q^2 < \mu^2$)  where $\mu$ is the resolution
defining the "resolvable" radiation,  equation (\ref {bfkl})
can be rearranged into the following "folded" form:
$$
f(x,Q_t^2)=\hat f^0(x,Q_t^2)+
$$
\begin{equation}
 \bar \alpha_s
\int_x^1{dx^{\prime}\over
x^{\prime}} \int {d^2 q\over \pi q^2} \Theta
(q^2-\mu^2)\Delta_R({x\over x^{\prime}},Q_t^2)
{Q_t^2 \over (\mbox{\boldmath $q$}+
\mbox{\boldmath $Q_t$})^2}
f(x^{\prime},(\mbox{\boldmath $q$}+
\mbox{\boldmath $Q_t$})^2) +O(\mu^2/Q_t^2)
\label{bfklr}
\end{equation}
where
\begin{equation}
\Delta_R(z,Q_t^2)=z^{\bar \alpha_s ln (Q_t^2/ \mu^2)}=
exp\left(-\bar \alpha_s\int_z^1{dz^{\prime}
\over z^{\prime}}\int_{\mu^2}^{Q_t^2}{dq^2\over q^2}\right)
\label{deltar}
\end{equation}
and
\begin{equation}
\hat f^0(x,Q_t^2)= \int_x^1{dx^{\prime}\over
x^{\prime}} \Delta_R({x\over x^{\prime}},Q_t^2)
{df^0(x^{\prime},Q^2_t))\over dln(1/x^{\prime})}
\label{hatf0}
\end{equation}
Equation (\ref{bfklr}) sums  the ladder diagrams
with  reggeized gluon exchange along the chain  with the gluon
trajectory $\alpha_G(Q_t^2) = 1-{\bar \alpha_s\over 2} ln(Q^2_t/\mu^2)$.\\

For the fixed coupling case  eq.(\ref{bfkl}) can be solved
analytically and the leading behaviour of its solution
at small $x$ is given by the
following expression:
\begin{equation}
f(x,Q_t^2) \sim (Q_t^2)^{{1\over 2}} {x^{-\lambda_{BFKL}}\over
\sqrt{ln({1\over x})}} exp\left(-{ln^2(Q_t^2/\bar Q^2)\over 2 \lambda^"
ln(1/x)} \right)
\end{equation}
\label{bfkls}
with
\begin{equation}
\lambda_{BFKL}=4 ln(2) \bar \alpha_s
\label{pombfkl}
\end{equation}
\begin{equation}
\lambda^"=\bar \alpha_s 28 \zeta(3)
\label{diff}
\end{equation}
where the Riemann zeta function $\zeta(3) \approx 1.202$.  The
parameter $\bar Q$ is of nonperturbative origin.\\

The quantity $1+ \lambda_{BFKL}$ is equal to the intercept of the so
called BFKL pomeron. Its potentially large magnitude ($\sim 1.5$)
should be contrasted with the intercept $\alpha_{soft} \approx 1.08$
of the (effective) "soft" pomeron which has been determined
from the phenomenological analysis of the high energy behaviour
of hadronic and photoproduction total cross-sections \cite{DOLA}.\\

In practice one introduces the running coupling $\bar \alpha_s(Q_t^2)$
in the BFKL equation (\ref{bfkl}). This requires introduction of the infrared
cut-off that would prevent entering the infrared region where the
coupling becomes large. The effective intercept $\lambda_{BFKL}$ found
by numerically solving the equation depends
on the magnitude of this cut-off \cite{KMS2}.\\

The structure functions $F_{2,L}(x,Q^2)$ are  described  at small $x$
by the
diagram of Fig.1 which gives the following relation between
the structure functions and the unintegrated distribution $f$:
\begin{equation}
F_{2,L}(x,Q^2)=\int_x^1{dx^{\prime}\over x^{\prime}}\int
{dQ_t^2\over Q_t^2}F^{box}_{2,L}(
x^{\prime},Q_t^2,Q^2)f({x\over x^{\prime}},Q_t^2).
\label{ktfac}
\end{equation}
The functions  $F^{box}_{2,L}(x^{\prime},Q_t^2,Q^2)$ may be regarded
as  the
structure
functions of the off-shell gluons with  virtuality
$Q_t^2$.
They are described by the quark box (and crossed box)
diagram contributions  to the
photon-gluon interaction in the upper part of the diagram of Fig. 1.
The small $x$ behaviour of the structure functions reflects the small
$z$ ($z = x/x^{\prime}$) behaviour of the gluon distribution $f(z,Q_t^2)$.\\

The equation (\ref{ktfac}) is an example of the "$k_t$ factorization theorem"
which relates  measurable quantities (like DIS structure functions) to
the convolution in both longitudinal as well as in transverse momenta of the
universal gluon distribution $f(z,Q_t^2)$ with the cross-section
(or structure function) describing the interaction of the "off-shell" gluon
with the hard probe \cite{KTFAC}.  The $k_t$ factorization theorem is a basic
tool for
calculating  observable quantities in the small $x$ region in terms of the
(unintegrated) gluon distribution $f$ which is the solution of the BFKL
equation.\\

A more general treatment of the gluon ladder is  provided by
the CCFM equation based on angular ordering along the gluon chain
\cite{CCFM,KMS1}.
This equation embodies both the BFKL equation at small $x$ and the
conventional Altarelli-Parisi evolution at large $x$.
The unintegrated gluon distribution $f$  now acquires
dependence upon an additional scale $Q$ which specifies
the maximal angle of gluon
emission.
The CCFM equation has a form analogous to that of the "folded" BFKL equation
(\ref{bfklr}):
$$
f(x,Q_t^2,Q^2)=\hat f^0(x,Q_t^2,Q^2)+ $$
\begin{equation}
\bar \alpha_s
\int_x^1{dx^{\prime}\over
x^{\prime}} \int {d^2 q\over \pi q^2} \Theta
(Q-qx/x^{\prime})\Delta_R({x\over x^{\prime}},Q_t^2,q^2)
{Q_t^2 \over (\mbox{\boldmath $q$}+
\mbox{\boldmath $Q_t$})^2}
f(x^{\prime},(\mbox{\boldmath $q$}+
\mbox{\boldmath $Q_t$})^2,q^2))
\label{ccfm}
\end{equation}
where the theta function $\Theta(Q-qx/x^{\prime})$ reflects the angular
ordering constraint on the emitted gluon.
The "non-Sudakov" form-factor $\Delta_R
(z,Q_t^2,q^2  )$ is now given by the following formula:
\begin{equation}
\Delta_R(z,Q_t^2,q^2)=exp\left[-\bar \alpha_s\int_z^1 {dz^{\prime}
\over z^{\prime}} \int {dq^{\prime 2}
\over q^{\prime 2}}\Theta (q^{\prime 2}-(qz^{\prime})^2)
\Theta (Q_t^2-q^{\prime 2})\right]
\label{ns}
\end{equation}
Eq.(\ref{ccfm}) still contains only the singular term of the
$g \rightarrow gg$ splitting function at small
 $z$ yet its generalization which would
include
the remaining parts of this vertex (as well as quarks) is possible.\\

In Fig. 2 we show the results for the structure function $F_2$ calculated
from the $k_t$ factorization theorem with the function $f$ obtained from
the CCFM equation \cite{CCFMF2}.
We confront these predictions with the most recent data
from the H1 and ZEUS collaborations at HERA \cite{H1,ZEUS} as well as
with the results
of the analysis which was based on the Altarelli-Parisi equation alone
without the small $x$ resummation effects being included in the formalism
\cite{MRS,GRV}.
In the latter case the singular small $x$ behaviour of the gluon
and sea quark distributions
 has to be introduced in a parametrization of the starting
distributions at the moderately large reference scale $Q^2=Q_0^2$
 (i.e. $Q_0^2 \approx 4GeV^2$ or so) \cite{MRS}.  One can also
generate the singular behaviour dynamically starting from the
non-singular "valence-like" parton distributions at some very low
scale $\mu_0^2=0.35GeV^2$ \cite{GRV}. In the latter case the gluon and sea
quark
distributions exhibit the following "double logarithmic behaviour"
\begin{equation}
xg(x,Q^2) \sim exp\left(2\sqrt{\xi(Q^2)ln({1\over x})}\right)
\label{dl}
\end{equation}
where the evolution length $\xi(Q^2)$ is defined as below:
\begin{equation}
\xi(Q^2)=\int_{\mu_0^2}^{Q^2} {dq^2\over q^2}\bar  \alpha_s(q^2) \sim
log \left({log({Q^2\over \Lambda^2})\over log({\mu_0^2\over \Lambda^2})
}\right)
\label{evl}
\end{equation}
For very small values of the scale $\mu_0^2$ the evolution length $\xi(Q^2)$
can become large for moderate and large values of $Q^2$ and the "double
logarithmic" behaviour (\ref{dl}) is within the limited region of $x$,
similar to that corresponding to the power like increase of the type
$x^{-\lambda}$, $\lambda \approx 0.3$.  This explains similarity between
the theoretical curves presented in Fig.2.
\medskip\medskip\medskip
\section*{4. Small $x$ behaviour of the nonsignlet unpolarized and polarized
structure functions}
The discussion presented in the previous Section concerned the small $x$
behaviour of the singlet structure function which was driven by the gluon
through the $g \rightarrow q \bar q$ transition.  The increase of the
gluon distribution in the small $x$ limit implies similar increase
of the structure function $F_2(x,Q^2)$.  It turns out to be stronger than
the increase that would follow from the "soft" pomeron exchange with its
relatively low intercept $\alpha_{soft} \approx 1.08$. \\

The gluons of course decouple from the non-singlet channel and the
mechanism of generating the small $x$ behaviour in this case is different.\\

The simple Regge pole exchange model predicts in this case that
\begin{equation}
F_2^{NS}(x,Q^2)=F_2^{p}(x,Q^2)-F_2^n(x,Q^2) \sim x^{1-\alpha_{A_2}(0)}
\label{a2}
\end{equation}
where $\alpha_{A_2}(0)$ is the intercept of the $A_2$ Regge
trajectory.  For $\alpha_{A_2}(0) \approx 1/2$ this behaviour is stable
against the leading order QCD evolution.  This follows from the fact that
the leading singularity of the moment $\gamma_{qq}(\omega)$ of the
splitting function $P_{qq}(z)$:
\begin{equation}
\gamma(\omega)=\int_0^1 {dz\over z} z^{\omega}P_{qq}(z)
\label{gqq}
\end{equation}
is located at $\omega=0$ and so
 the (nonperturbative) $A_2$ Regge pole at $\omega=\alpha_{A_2}(0) \approx
1/2$  remains the leading
singularity controlling the small $x$ behaviour of the non-singlet
structure function.\\

The novel feature of the non-singlet channel is the appearence of the
{\bf double} logarithmic  terms i.e. powers of
$\alpha_s ln^2(1/x)$ at each order of the perturbative
expansion.  These double logarithmic terms are generated by the
ladder diagrams with  quark (antiquark) exchange along the chain.
The ladder diagrams can acquire corrections from the "bremsstrahlung"
contributions \cite{KL,BER}
which do not vanish for the polarized structure function
$g_1^{NS}(x,Q^2)$ \cite{BER}.\\

In the approximation when the leading double logarithmic terms
are generated by ladder diagrams illustrated in Fig. 3 the
unintegrated non-singlet quark distribution $f_q^{NS}(x,k^2_t)$
($q^{NS}=u+\bar u - d -\bar d$) satisfies
the following integral equation  :
\begin{equation}
f_q^{NS}(x,Q_t^2)=f_{q0}^{NS}(x,Q_t^2)+ \tilde \alpha_s
\int_x^1{dz\over z}\int_{Q_0^2}^{{Q_t^2\over z}} {dQ_t^{\prime 2}\over
Q_t^{\prime 2}}f_q^{NS}({x\over z},Q_t^{\prime 2})
\label{dleq}
\end{equation}
where
\begin{equation}
\tilde \alpha_s = {2 \over 3 \pi} \alpha_s
\label{atil}
\end{equation}
and $Q_0^2$ is the infrared cut-off parameter.
The unintegrated distribution $f_q^{NS}(x,Q_t^2)$ is, as usual, related
in the following way to the scale dependent (nonsinglet) quark distribution
$q^{NS}(x,Q^2)$:
\begin{equation}
q^{NS}(x,Q^2)=\int^{Q^2}{dQ_t^2\over Q_t^2}f_q^{NS}(x,Q_t^2).
\label{intd}
\end{equation}
The upper limit $Q_t^2/z$ in the integral equation (\ref{dleq}) follows
from the
requirement that the virtuality of the quark at the end of the chain
is dominated by $Q_t^2$. A possible non-perturbative $A_2$ reggeon contribution
has to be introduced in the driving tem i.e.
\begin{equation}
f_{q0}^{NS}(x,Q_t^2) \sim x^{-\alpha_{A_2}(0)}
\label{a2driv}
\end{equation}
at small $x$.\\

Equation (\ref{dleq}) implies the following equation
for the moment function $\bar f_q^{NS}(\omega,Q_t^2)$
\begin{equation}
\bar f_q^{NS}(\omega,Q_t^2)=\bar f_{q0}^{NS}(\omega,Q_t^2)+
{\tilde \alpha_s \over \omega} \left[\int_{Q_0^2}^{Q_t^2}
{dQ_t^{\prime 2}\over
Q_t^{\prime 2}}\bar f_q^{NS}(\omega,Q_t^{\prime 2})+
\int_{Q_t^2}^{\infty} {dQ_t^{\prime 2}\over
Q_t^{\prime 2}}\left({Q_t^2 \over Q_t^{\prime 2}}\right)^{\omega}
\bar f_q^{NS}(\omega,Q_t^{\prime 2})\right]
\label{dleqm}
\end{equation}
Equation (\ref{dleqm}) follows from (\ref{dleq}) after taking  into
account the following relation:
\begin{equation}
\int_0^1{dz\over z}z^{\omega}\Theta \left({Q^2_t\over Q^{\prime 2}_t} - z
\right)=
{1\over \omega}\left[\Theta(Q_t^2-Q^{\prime 2}_t)+
\left({Q^2_t\over Q^{\prime 2}_t}\right)^
{\omega}\Theta (Q^{\prime 2}_t-Q^{2}_t)\right].
\label{theta}
\end{equation}
For fixed coupling $\tilde \alpha_s$  equation (\ref{dleqm})
 can be solved analytically.
Assuming for simplicity that the inhomogeneous term is independent
of $Q_t^2$ (i.e. that $\bar f_{q0}^{NS}(\omega,Q_t^2) = C(\omega)$ )
we get the following solution of  eq.(\ref{dleqm}):
\begin{equation}
\bar f_q^{NS}(\omega,Q_t^2)=C(\omega)R(\tilde \alpha_s,  \omega)
\left({Q_t^2\over Q_0^2}\right)^{ \gamma^{-}(\tilde \alpha_s,  \omega)}
\label{solm}
\end{equation}
where
\begin{equation}
\gamma^{-}(\tilde \alpha_s, \omega) = {\omega - \sqrt{\omega^2 - 4 \tilde
 \alpha_s}\over 2}
\label{anomd}
\end{equation}
and
\begin{equation}
R(\tilde \alpha_s,  \omega)= {\omega \gamma^{-}(\tilde \alpha_s, \omega)\over
\tilde \alpha_s}.
\label{r}
\end{equation}
Equation (\ref{anomd}) defines the anomalous dimension of the
 moment of the non-singlet quark distribution in which
 the double logarithmic $ln(1/x)$ terms i.e. the powers of ${\alpha_s \over
\omega^2}$ have been resummed to all orders.  It can be seen from (\ref{anomd})
that this anomalous dimension has a (square root) branch point singularity
at $\omega=
\bar \omega$
\begin{equation}
\bar \omega= 2 \sqrt{\tilde \alpha_s}
\label{barom}
\end{equation}
This singularity will of course be also present in the moment function $
\bar f_q^{NS}(\omega,Q_t^2)$ itself. It should be noted that in contrast to the
BFKL singularity whose position above unity was proportional to $\alpha_s$,
$\bar \omega$ is proportional to $\sqrt{\alpha_s}$ - this being the
straightforward consequence of the fact that  equation (\ref{dleqm})
sums double logarithmic terms $({\alpha_s\over \omega^2})^n$.
This singularity gives the following contribution to the
non-singlet quark distribution $f_q^{NS}(x,Q_t^2)$ at small
$x$:
\begin{equation}
f_q^{NS}(x,Q_t^2) \sim {x^{-\bar \omega}\over ln^{3/2}(1/x)}
\label{smxns}
\end{equation}
For small values of the QCD  coupling this contribution remains non - leading
in comparison to the contribution of the $A_2$ Regge pole.\\

 As  has been  mentioned above the
corresponding integral equation which resums the double logarithmic
terms  in the spin dependent quark distributions is more
complicated than the simple ladder equation (\ref{dleq})
due to non-vanishing contributions coming from bremsstrahlung diagrams.
It may however be shown that , at least as far as the non-singlet
structure function is concerned, these contributions give only relatively small
correction to $\bar \omega$.  In what follows we shall therefore limit
ourselves to the simple ladder equation (\ref{dleq}) assuming
that it  will describe the spin dependent parton densities as well.
The inhomogeneous term will now however be different
and will contain the $A_1$ reggeon contribution.
We will limit ourselves to the non-singlet structure function
$g_1^{NS}(x,Q^2) = g_1^p(x,Q^2)-g_1^n(x,Q^2) = \Delta q^{NS}/3$ where
\begin{equation}
\Delta q^{NS}(x,Q^2)=\Delta u(x,Q^2)+\bar \Delta u(x,Q^2) -
\Delta d(x,Q^2) - \bar \Delta d(x,Q^2)
\label{gns}
\end{equation}
where $\Delta u, \Delta d$ and $\bar \Delta u, \bar \Delta u$ denote
the corresponding spin dependent quark (antiquark) distributions.
We will present
an  estimate of the effective slope of the non-singlet distributions after
numerically solving  equation (\ref{dleq}) taking into account
the asymptotic freedom corrections i.e. allowing the coupling constant
$\alpha_s$ to run.\\

The main interest in applying the QCD evolution equations to study
the spin structure function is that the naive Regge pole expectations based
on the exchange of  low-lying Regge trajectories become unstable against
the QCD perturbative "corrections".  The relevant reggeon which contributes
to $g_1^{NS}(x,Q^2)$ is the  $A_1$  exchange which
is expected to have a very low intercept $\alpha_{A_1}(0) \le 0$.
The perturbative singularity generated by the double logarithmic
$ln(1/x)$ resummation can therefore become much more important
than in the case of the unpolarized case when it is hidden behind
  leading $A_2$ exchange contribution.  Even if we restrict ourselves
to the leading order QCD evolution
\cite{GGR,BFR}
then the non-singular $x^{-\alpha_{A_1}(0)}$ behaviour
(with $\alpha_{A_1}(0) \le 0$ ) becomes unstable as well
and the polarized quark
densities acquire singular behaviour:
\begin{equation}
\Delta q^{NS}(x,Q^2) \sim exp(2 \sqrt{\xi^{NS}(Q^2)ln(1/x)})
\label{dqdl}
\end{equation}
where
\begin{equation}
\xi^{NS}(Q^2)=\int^{Q^2} {dq^2 \over q^2} \tilde \alpha_s(q^2)
\label{xins}
\end{equation}
This follows from the fact that $\Delta P_{qq}(z)=P_{qq}(z)$
where $P_{qq}(z)$  and  $\Delta P_{qq}(z)$ are the splitting functions
describing the evolution of spin independent and spin dependent quark
distributions respectively and from the fact that $P_{qq}(z) \rightarrow
const$ as $z \rightarrow 0$.\\

The introduction of the running coupling effects in  equation (\ref{dleqm})
turns the branch point singularity into the series of poles which accumulate
at $\omega=0$.  If one makes the substitution $\tilde \alpha_s \rightarrow
\tilde \alpha_s(Q_t^2)$ then the corresponding equation for the moment
can still be solved analytically by reducing it to  Kummer's
differential equation with the boundary conditions fixed
 by the requirement that the solution should
match  the perturbative expansion of the original
integral equation \cite{JK2}.
It has however been argued that the theoretically more justified
introduction of the running coupling is through the substitution
$\tilde \alpha_s \rightarrow \tilde \alpha_s(Q_t^2/z)$ under the integrand
on the rhs. of  eq. (\ref{dleq}) \cite{EMR}.  We  perform a numerical
analysis with both prescriptions and  estimate the effective
slope controlling the small $x$ behaviour of the solution.
The two prescriptions of introducing the running coupling effects
into the double $ln(1/x)$ resummation lead to the following
integral equation(s) for $\Delta f_q^{NS}(x,Q_t^2)$
\begin{equation}
\Delta f_q^{NS}(x,Q_t^2)=\Delta f_{q0}^{NS}(x,Q_t^2)+ \tilde \alpha_s(Q_t^2)
\int_x^1{dz\over z}\int_{Q_0^2}^{{Q_t^2\over z}} {dQ_t^{\prime 2}\over
Q_t^{\prime 2}}\Delta f_q^{NS}({x\over z},Q_t^{\prime 2})
\label{eqdlbz}
\end{equation}
or
\begin{equation}
\Delta f_q^{NS}(x,Q_t^2)=\Delta f_{q0}^{NS}(x,Q_t^2)+
\int_x^1{dz\over z}\tilde \alpha_s(Q_t^2/z)
\int_{Q_0^2}^{{Q_t^2\over z}} {dQ_t^{\prime 2}\over
Q_t^{\prime 2}}\Delta f_q^{NS}({x\over z},Q_t^{\prime 2}).
\label{eqdlzz}
\end{equation}
We  solve these equations numerically,  suitably adapting the method
developed in \cite{JKBFKL}. We assume for simplicity that the inhomogeneous
term $ \Delta f_{q0}^{NS}(x,Q_t^2)$ is independent of both $x$ and of $Q_t^2$.
The flat $x$ dependence of the inhomogeneous term corresponds to the
assumption that $\alpha_{A_1}(0)=0$.
We assume the LO parametrization of the running coupling
with four flavours and set $\Lambda=0.2GeV$. The cut-off
parameter $Q_0^2$ was set equal to $Q_0^2=1GeV^2$.
\medskip
In Fig.4 we show the effective slope $\lambda(x,Q_t^2)$ for
$Q_t^2=10GeV^2$,
\begin{equation}
\lambda(x,Q_t^2)={dln \Delta f_q^{NS}(x,Q_t^2)\over d ln(1/x)}
\label{slope}
\end{equation}
for the solutions of  equation (\ref{eqdlbz})
and of  equation (\ref{eqdlzz}).
We also show the effective slope corresponding to the approximation
of retaining only  the first integral on the right hand side
of  eq. (\ref{eqdlbz}).
In this approximation one sums the single logarithmic ln$(1/x)$ terms
accompanied by the leading powers of $\xi^{NS}(Q_t^2)$. The small
$x$ behaviour is then asymptotically given  by  eq.(\ref{xins}).
We have also solved  eq.(\ref{eqdlzz}) assuming that the driving term is
given by the $A_2$ reggeon contribution (see eq.(\ref{a2driv})). The
resulting slope of the solution is also shown in Fig. 4.
It can be seen that the second prescription (i.e
$\alpha_s \rightarrow \alpha_s(Q_t^2/z)$) for
introducing the running coupling effects makes the effective slope
smaller than  in the case when
one makes the  substitution $\alpha_s \rightarrow \alpha_s(Q_t^2)$.
The perturbative Regge singularities are  not very important
for the unpolarized structure functions and the (input) $x^{-1/2}$ behaviour
is not altered substantially by the perturbative Regge singularity.
The double logarithmic resummation is  however very important for generating
the singular small $x$ behaviour of the polarized  structure functions.
The results of our estimate suggest that a reasonable extrapolation
of the (non-singlet) polarized quark densities would be to assume an
$x^{-\lambda}$ behaviour with $\lambda \approx 0.2$ \footnote{
The effective slope $\lambda$ turns out to depend weakly upon the
infrared cut-off parameter $Q_0^2$.  If we set $Q_0^2=\Lambda^2$ then
$\lambda \rightarrow {8\over (33-2N_f)}$ = 0.32. I thank Misha Ryskin
for pointing out this fact to me.}.  Similar
(or more singular) extrapolations of the spin-dependent quark
distributions towards the small $x$ region have
 been assumed in  several recent parametrizations of parton densities
\cite{BS,GRVOG,GS,BV}. \\

\section*{5. Summary and conclusions}
In this paper we have briefly summarized the theoretical QCD expectations
concerning the small $x$ behaviour of the deep inelastic scattering
structure functions in both unpolarized and polarized deep inelastic
scattering. In the latter case we have  for simplicity focussed on the
non-singlet structure functions which, at small $x$ , can be (approximately)
described by the ladder diagrams with the quark (antiquark) exchange.
We have solved the corresponding integral equation taking into account
 asymptotic freedom effects and estimated the effective slope
controlling the small $x$ behaviour of the non-singlet structure function.
The perturbative QCD effects become significantly amplified for the
singlet spin structure function due to the mixing with the gluons.
The simple ladder equation may not however be  applicable
for an accurate description of the double logarithmic terms in
the polarized gluon distribution $\Delta G$.\\

\section*{Acknowledgments}
I thank Alan Martin and Peter Sutton for most enjoyable research collaboration
on the problems presented in this paper.  I thank them
as well as Jochen Bartels, Misha Ryskin and Andreas Sch\"afer
for illuminating discussions. I am grateful to Grey College and Physics
Department of the University of Durham for their warm hospitality.
This research has been supported in part by
 the Polish State Committee for Scientific Research grant 2 P302 062 04 and
the EU under contracts n0. CHRX-CT92-0004/CT93-357.\\

\vspace{10mm}
{\Large {\bf Figure captions}}
\begin{enumerate}
\item
Diagrammatic representation of the $k_t$ factorization formula (\ref{ktfac}).
\item
A comparison of the HERA measurements of $F_2$ \cite{H1,ZEUS} with
the predictions based on the $k_t$ factorization formula (\ref{ktfac})
using for the unintegrated gluon distributions $f$ the solutions of the
CCFM equation (\ref{ccfm}) (continuous curve) and of the
approximate form of this
equation corresponding to setting
$\Theta(Q-q)$ in place of $\Theta(Q-qx/x^{\prime})$ and
$\Delta_R=1$ (dotted curve).
 We also show the values of $F_2$ obtained from
collinear factorization using the MRS(A$^{\prime})$ \cite{MRS} and
GRV \cite{GRV} partons (the figure is taken from  ref. \cite{CCFMF2}).
\item
The ladder diagram with quark (antiquark) exchange along the chain.
\item
The effective slope $\lambda(x,Q^2_t)$ defined by the formula
(\ref{slope}) of the solutions of the equations (\ref{eqdlbz})
(dashed curve) and
(\ref{eqdlzz}) (solid curve). The dashed-dotted curve
corresponds to  the approximation of  eq. (\ref{eqdlbz})
in which only  the first integral on  right hand side is retained.
The dotted curve represents the slope of the solution
of  eq. (\ref{eqdlzz}) in which the inhomogeneous term is set
proportional
to $x^{-1/2}$.  The slopes are plotted as  functions of $x$ for
fixed $Q_t^2=10GeV^2$.
\end{enumerate}
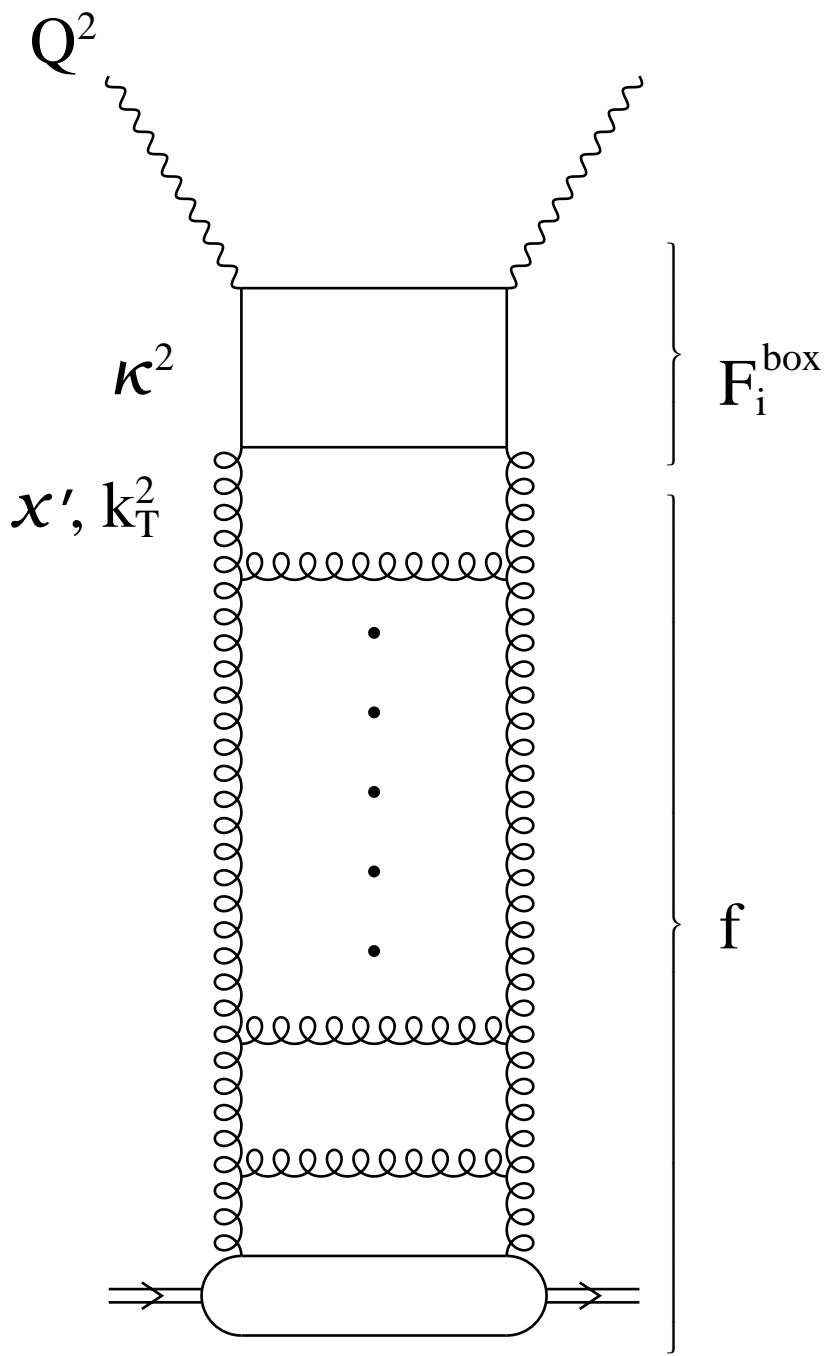
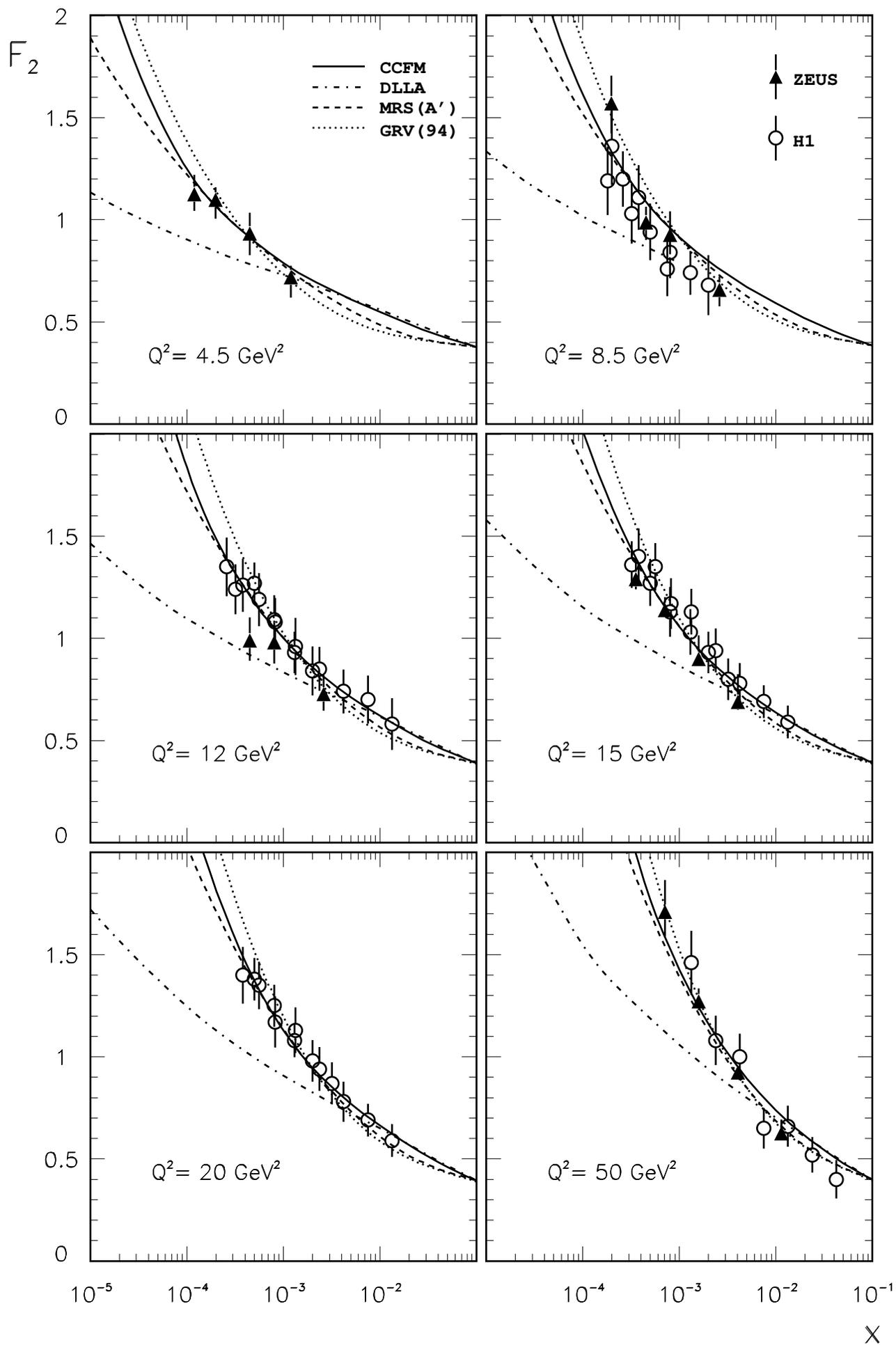
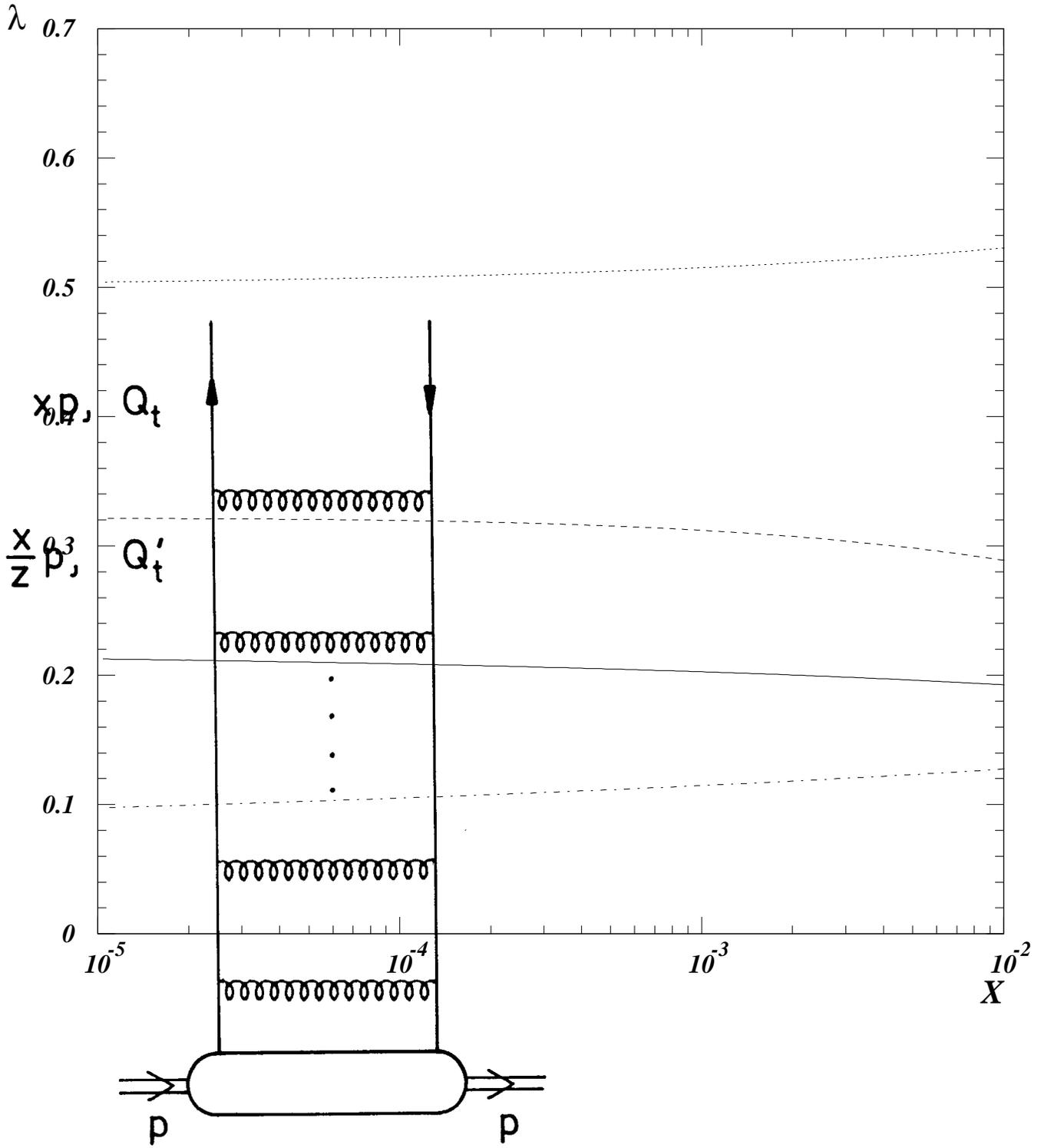
\end{document}